# Enhancement of CASSI by a zero-order image employing a single detector

**J. Hlubuček\*, J. Lukeš, J. Václavík, AND K. Žídek**

*Regional Center for Special Optics and Optoelectronic Systems (TOPTEC), Institute of Plasma Physics, Czech Academy of Sciences v.v.i., Sobotecká 1660, 511 00 Turnov, Czech Republic*
*\*hlubucek@ipp.cas.cz*

**Abstract:** Coded aperture snapshot spectral imaging (CASSI) makes it possible to recover 3D hyperspectral data from a single 2D image. However, the reconstruction problem is severely underdetermined and efforts to improve the compression ratio typically make the imaging system more complex and cause a significant loss of incoming light intensity. In this paper, we propose a novel approach to CASSI which enables capturing both spectrally sheared and integrated image of a scene with a single camera. We performed hyperspectral imaging of three different testing scenes in the spectral range of 500-900 nm. We demonstrate the prominent effect of using the non-diffracted image on the reconstruction of data from our camera. The use of the spectrally integrated image improves the reconstruction quality and we observed an approx. fivefold reduction in reconstruction time.

## 1. Introduction

Hyperspectral imaging (HSI) instrumentation is essential for many applications ranging from scientific research, such as volcanology [1] or imaging the chiralities of single nanotubes [2], to practical problems including food analysis and safety inspection [3, 4], medical imaging [5], quality control [6], forensic sciences [7, 8], or art conservation [9].

Besides standard methods, commonly used to acquire a HS datacube, such as whiskbroom, pushbroom, and plane scanning, a range of new techniques have been developed with the vision to create a single-snapshot HSI, which can be operated with a high frame rate and does not require any movable part [10]. One of the methods is CASSI (coded aperture snapshot spectral imaging), based on compressed sensing [11, 12].

CASSI can outperform the standard techniques mainly in the length of the acquisition time since it captures the whole datacube in one instance, i.e., a snapshot, eliminating the need for scanning. This makes the system highly robust. At the same time, the single-frame CASSI system has certain limitations, including image quality, compression ratio, and the time needed for the HS datacube reconstruction, since the reconstruction problem is severely underdetermined.

It is possible to improve the reconstruction quality of CASSI, for instance, by optimizing a coded aperture [13,14], utilizing multiple camera shots [15-17], or using a higher-order image reconstruction [18]. On the other hand, refining the method often brings in certain limitations. Multi-frame CASSI loses the advantages of using a single snapshot, while more complex models for the detector description slow down the reconstruction process. Another promising way to boost the performance of CASSI is to capture a non-diffracted image which aids in the reconstruction. However, this approach normally requires splitting an incoming beam and employing two cameras [19-23], which makes the CASSI system inconveniently complex and causes a loss in the light intensity, which can reach as much as 50% [23].

Another limitation of the CASSI method consists in the size of the measurable spectral range. The spectral reconstruction can be highly improved by identifying key spectral features in the spectrum for specific applications [24]. This is, however, not our case, as we aim at a reconstruction of an arbitrary spectral shape, including spectrally flat broadband sources.

Acquisition of a broader bandwidth decreases the compression ratio, which lowers the quality of the retrieved hyperspectral information. Therefore, the above-mentioned upgrades of CASSI typically aim at increasing the compression ratio along with capturing a narrow spectral range. At the same time, the CASSI reconstruction assumes an ideal image for each wavelength, which brings in the necessity to highly reduce optical aberrations of the CASSI system in the case of spectrally broad light. This leads to complex optical systems limited in their spectral range. Hence, there is a trade-off between the ability to carry out broadband HSI and the complexity of the setup. This is even more prominent in the infrared (IR) spectral range, where the construction of complex systems is costly and their precise alignment is a challenging task.

In this article, we present a robust concentric HS camera based predominantly on off-the-shelf optics, which can be used for CASSI HSI. In contrast to previous reports, we aim at obtaining HSI in a broad spectral range between 500-900 nm covered by 123 spectral frames. In combination with the simplicity of the camera, the broad spectral range leads to the presence of aberrations in the system. This camera serves as a model system for the perspective of the IR CASSI imaging, where the acquisition of a broad spectral range is needed to capture and distinguish between different chemical agents in the IR region.

However, the uniqueness of our HS camera lies in the design of the dispersive elements which are able to attain both a non-diffracted image and first-order diffraction with a single detector. We demonstrate that by using a zero-order image of a diffraction grating, we can highly improve the reconstruction quality of the system in spite of the aberrations present. Moreover, owing to the camera construction used, we utilize the light intensity which is otherwise dumped in the other grating-based CASSI systems [23].

By providing measurements of three testing scenes, we show that the use of zero-order diffraction is indispensable for the aberrated system in order to attain spatial quality of HS datacube reconstruction. This is particularly prominent for scenes of spectrally broad light. We compare the use of the zero-order in the calculation of an initial guess in the iterative reconstruction, as well as in the reconstruction itself. The presented concept can serve as an efficient approach to improving reconstruction in CASSI systems suffering from aberrations and low compression ratio.

## 2. Experimental setup

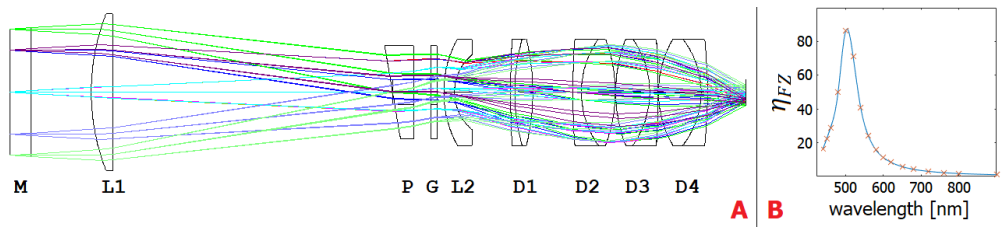

Figure 1. (A) Scheme of our system described in the text, (B) Spectral dependence of a relative intensity between first-order image and zero-order image $\eta_{FZ}$.

The used hyperspectral camera, depicted in Fig. 1A, was built based on off-the-shelf optics except for elements L2 and P, which were manufactured at our facilities. Its main features are a high numerical aperture (NA~0.35) and a telecentric object (mask) space. In the scheme, L denotes plano-convex lenses, D denotes doublets, M is a random mask, P is a prism, and G is a grating. A detailed description of the system with a list of all its elements can be found in [25]. A total of six optical elements available from optics catalogs and a custom-made lens (L2) and prism (P) were used for the construction. A combination of the transmission grating (G, Thorlabs, 300 lines/mm) and the custom-made prism (P, SF11 optical glass) allows for a concentric construction of the camera, which is beneficial for calibration, and it also enables simple mechanical housing into a single tube. Mask M was a binary pattern, which was prepared via photolithography on a BK7 substrate with a thin chromium layer. It has 64x64 pixels and a side length of 13.55 mm. The resulting image, which consists of both first-order (FO) and zero-order (ZO) diffraction, was detected by using a Manta G-507 camera (Sony IMX264, resolution 2464×2056).

Due to the different spectral response of the optical system for the FO and the ZO, we characterized the relative intensity between the FO and the ZO intensities, which we denote as $\eta_{FZ}$. The intensity ratio, affected dominantly by the grating response, is depicted in Fig. 1B. The spectral efficiency of the FO vs. the ZO was employed in the calculations to reliably reproduce the detector image in Eq. (2). The monochromatic light for spectral calibration was obtained using a monochromator (Chromex 250 IS) in combination with a broadband quartz tungsten-halogen lamp (Thorlabs).

For the sake of the testing experiments, the testing scenes described below were imaged on the mask M by a single thin lens combined with a cut-off filter OG-515, which restricted the measured spectral range below 500 nm, as we explain below.

## 3. Data processing and reconstruction

HS datacube reconstruction requires a transfer of the captured detector image with a high resolution (2464×2056 pixel) into an image of the FO and the ZO corresponding to the resolution of the random mask (64×64 pixel). First, the detector image is cropped and resized to match the pixel size of the random mask. The cropping employs calibration with a diffused monochromatic light (Nd:YAG laser, 532 nm). The crude cropping is based on aim pointers. These are transparent pixels located in the proximity of the mask, which can be identified in the dark detector area. Owing to the narrow spectrum of the calibration laser, the image of the diffused laser light on the detector is an image of the random mask without any spectral shear. We determined the cropping range of both orders by searching for the best correlation between the image and the random mask.

Since the detector has a higher resolution than the random mask – one mask pixel corresponds to approximately eight pixels on the detector – it is necessary to resize the cropped image. For the sake of contrast improvement, we avoid the border pixels, which, in the sense of binary mask pattern, could be classified as 'gray'. For the zero-order image, the border pixels are avoided in both directions, while for the first-order one, the omission can be performed only in the direction of spectral shearing. The image, where the border pixels were nullified, is consequently rescaled into a 64x186 pixel FO image and a 64x64 pixel ZO image corresponding to the mask pixels.

The processed data are reconstructed using the TwIST algorithm [26] minimizing the expression:

$$f(D) = \frac{1}{2}\left\|I - \widehat{W}D\right\|^2 + \lambda \Phi(D) \qquad (1)$$

where $I$ is the detector output; $\widehat{W} = \hat{S}\widehat{M}\widehat{H}^{-1}$ is an operator describing the propagation of the incoming light through the system, including modulation by the random mask $\widehat{M}$ and the spectral slice placement $\hat{S}$ in the FO and ZO images. $D(x, y, \lambda)$ is the HS datacube, where each spectral frame is transformed by the Haar wavelet transform $\widehat{H}$. We use $l_1$-norm regularization $\Phi(D) = \sum|D|$ since the common scenes are sparse-like in the Haar wavelet basis. The regularization term is weighted by a coefficient $\lambda$, which can emphasize the sparsity of the reconstructed datacube.

The FO and the ZO can be included in the operator of spectral shearing $\hat{S}$ as:

$$\hat{S} = \sum_\lambda \left[\eta_{FZ}(\lambda)\hat{T}_1(\lambda) + \hat{T}_0\right] \qquad (2)$$

where $T_1(\lambda)$ is the wavelength-dependent translation of the image to the FO area and $T_0$ is the wavelength-independent translation of the image to the ZO area. The coefficient $\eta_{FZ}(\lambda)$ is the measured spectral efficiency of the FO vs. the ZO depicted in Figure 1B.

The TwIST algorithm uses two operators: (i) to transform the datacube to the detector image, and (ii) to transform the detector image into the datacube. These operators correspond, in the compressed sensing theory, to the sensing operator $\widehat{W}$ and to its transposition $\widehat{W}^T$, respectively. Since a matrix representation of $\widehat{W}$ and $\widehat{W}^T$ would be very large and unsuitable for fast reconstruction, we evaluate them as functions.

Since TwIST is an iterative algorithm, an important factor is the initial guess of the hyperspectral datacube. Therefore, the ZO image can be implemented not only in the TwIST reconstruction itself but also in the initial guess. It is highly favorable, in the sense of reconstruction time and quality, to make the initial guess as similar to the real datacube as possible.

An issue connected with the use of the ZO consists in the fact that the ZO image has blank pixels where the binary values of the random mask are equal to zero. We have overcome this by approximating these pixels by the mean value of their neighboring pixels. At the same time, the reconstruction needs to take into account that the ZO image is not evenly represented by all wavelengths.

The initial guess was created from the detector output $I$ for each wavelength of a spectral slice as follows:
1) We extracted the 64x64 pixel spectral slice $\Gamma(\lambda)$ of the HS datacube from the detector FO image, where the slice position corresponds to the selected wavelength $\lambda$. We multiplied the slice with the random mask: $\Gamma(\lambda) = \widehat{M}\widehat{T}_1^{-1}(\lambda)\,I$.
2) The spectral weight of the slice was calculated as a sum of all elements of the slice $\Gamma(\lambda)$: $w(\lambda) = \sum_{x,y} \Gamma(\lambda)$.
3) The ZO image $Z$ extracted from the detector was used to correct the spectral slice after normalization by its mean value $\bar{Z}$: $\tilde{\Gamma}(\lambda) = \frac{Z}{\bar{Z}}(\Gamma(\lambda) + Z)$.
4) The initial guess $G(\lambda)$ was obtained by treating the $\tilde{\Gamma}(\lambda)$ slice with total variation denoising $\widehat{N}$ corresponding to the Rudin-Osher-Fatemi denoising model and the denoised slice is multiplied by its spectral weight: $G(\lambda) = w(\lambda).\widehat{N}\Gamma(\lambda)$.

The resulting datacube guess G is finally rescaled by the ratio between the original detector image intensity and the detector image intensity obtained by applying the operator $\widehat{W}$ to the datacube guess.

## 4. Results and discussion

We carried out a set of experiments where we studied the hyperspectral datacube reconstruction from our broadband aberrated hyperspectral camera based on the CASSI method. As we described in the previous two sections, besides the standard CASSI method, where the image is modulated by a mask M and spectrally dispersed, the construction of our device makes it possible to capture also the ZO diffraction. Hence, we exploit a part of the light intensity that would otherwise be lost in a standard system and we use it to gain more information about the measured scene. The aim of the experiments was to reveal the effect of the information about the zero-order image on the datacube reconstruction.

In order to gain a quantitative evaluation of the reconstruction quality, we created artificial data faithfully simulating the real detected FO images as well as the ZO image by a careful analysis of the aberrations present in our system. Namely, we included the effect of wavelength-dependent: (i) vertical shift of spectral slices on the detector, and (ii) spectral slices acutance. The effects were simulated by varying the size and position of each spectral slice jointly with a wavelength-dependent Gaussian filter. The scale of aberrations was extracted based on the acquired images of a monochromatic source illuminating a mask (see Fig. 2a), as we discussed in the next paragraphs. The simulated data were calculated to follow the camera resolution and they were processed and reconstructed by using the exact same procedure as the experimental data.

The simulations allowed us to compare the reconstructed datacube with the ground-truth and, therefore, attain a quantitative measure of the reconstruction quality. We define the *difference Δ* between the ground-truth and the reconstructed artificial datacube by least squares, where we optimize scaling factor *c* to minimize the *difference* value:

$$\Delta = \min_c \left\{ \frac{1}{n} \sum_{i=1}^{n} (c.y_i - \bar{y}_i)^2 \right\}, \quad (3)$$

where residuals for i-th point are calculated as a difference between the original datacube value $y_i$ and reconstructed datacube value $\bar{y}_i$ and $n$ is the number of datacube voxels.

The camera was constructed with a primary restriction on the number of elements and their off-the-shelf availability and, at the same time, we use the camera on a broad spectral range of 500-900 nm. Due to these restraints, the resulting detector image is aberrated.

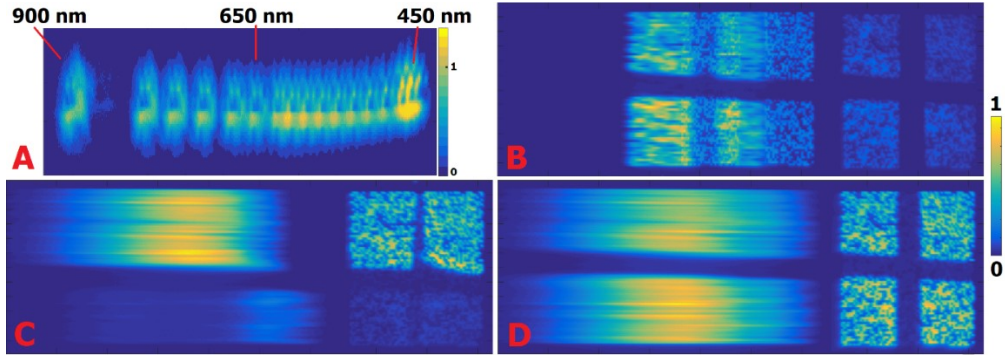

Figure 2. (A) FO images of a fixed spot on the random mask illuminated by a set of monochromatic lights with wavelengths ranging from 440 nm to 900 nm (superimposed normalized images). Differences in the spot vertical position, scaling, and sharpness emerge from aberrations in the FO image, image resolution: 80x1050 px, (B-D) scenes reconstructed in the article (normalized, colorbar on the right), image resolution 600x2260px: (B) quasi-monochromatic laser sources illuminating dark cross; (C) spectrally broad light transmitted through four color filters; (D) spectrally broad light illuminating dark cross.

We can visualize the aberrations present in the system (see Fig. 2A) by a superposition of detected images of a single spot on a mask illuminated by a set of quasi-monochromatic wavelengths. The detector was consequently illuminated in the spectral range of 440-900 nm, where the wavelength of the imaged spot decreases from left to right on the detector. The image for each wavelength was normalized before being added to the overall sum. Note that, compared to the other panels in Fig. 2, panel A is highly rescaled to demonstrate the aberrations.

For wavelengths around 450-500 nm (Figure 2A, on the right), the spot vertical position changes rapidly. This discrepancy is around one mask pixel, which makes the correct reconstruction impossible. Therefore, we used the OG-515 filter to block this problematic part of the spectrum. At the same time, you can see that the image acutance changes with wavelengths, and a sharp image is obtained only in the central part of the spectrum. This is another source of imperfections in the reconstruction.

Figure 2 B-D shows a detector output of three different scenes: (i) an opaque cross illuminated simultaneously by a green laser and a red diode (Scene A, Figure 2B); (ii) four color filters illuminated by a broadband light source (Scene B, Figure 2C); (iii) an opaque cross illuminated by a broadband light source (Scene C, Figure 2D). On the left-hand side in the respective pictures, you can see the first-order diffraction and on the right is the zero-order diffraction. In Figure 2B the intensity of the zero-order diffraction is very weak, which is caused by using only two wavelengths and by the spectral effectivity of the FO vs. the ZO. As you can see, the FO image of the green laser is basically an image of the random mask, since the laser spectral width is well below the spectral resolution of our system. On the contrary, the FO image of the red diode is a bit sheared due to the spectral width FWHM being 18.5 nm.

It is worth noting, that the spatial resolution of reconstructed scenes is restricted by the resolution of the used mask M and not by the detector. While the used photolithographic process allows fabrication of a much finer binary mask, the resolution is limited by the aberrations present in our system. For example, in the spectral range where the used light source is the most intense, i.e., 520-720 nm, the variance in the vertical shift of different images is less than ¼ of the mask pixel. This is still a feasible inaccuracy as we do not include in our calculations the border pixels between the lines of the mask (discussed in Chapter 3). However, the use of a finer mask, i.e. smaller pixels, would inevitably lead to wrongly encoded information on the detector, where the information from one mask line would leak into the neighboring ones.

To study the influence of the ZO in the reconstruction, it can be implemented in two ways: First, to improve the initial guess of the reconstruction, and secondly, to be included in the operator $\hat{W}$ in the TwIST.

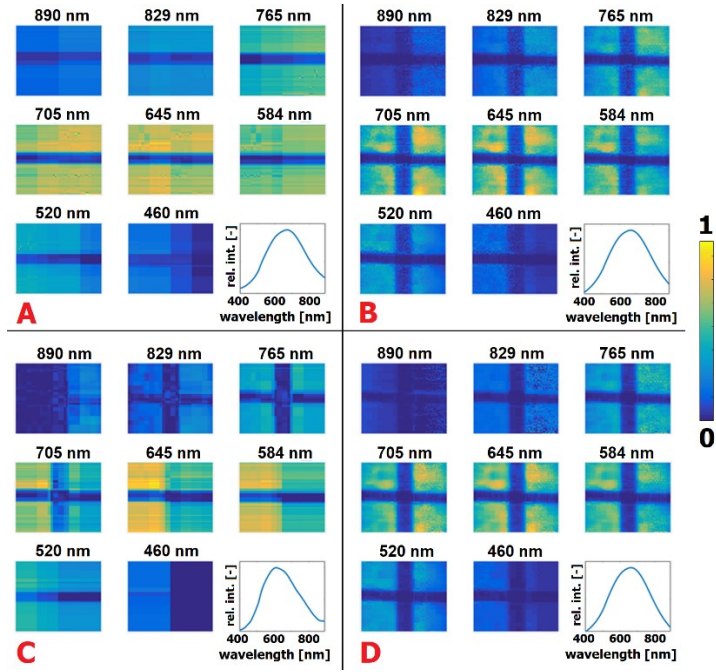

Figure 3. Reconstruction of the scene from Figure 2D; each selected spectral slice is normalized to the maximum datacube value, colorbar on the right; (A) Not using ZO, (B) using ZO in initial guess only, (C) using ZO in operator $\hat{W}$ only, (D) using ZO both in initial guess and operator $\hat{W}$.

Figure 3 depicts the effect of four different modes of (not) using the ZO: (panel A) standard CASSI reconstruction with no ZO information; (panel B) ZO-assisted initial guess calculation followed by a standard CASSI reconstruction avoiding the ZO inclusion; (panel C) initial guess omitting the ZO information while using the ZO in the TwIST reconstruction; and (panel D) using the ZO in both the initial guess estimate and the datacube reconstruction. All calculations employed the scene with an opaque cross illuminated by a broadband light source. In each panel, we present eight selected spectral slices of the datacube together with the total spectrum in the bottom right graph, which is a sum of all elements of each slice.

As one can see in Figure 3A, without the information about the ZO it was not possible to retrieve the vertical line of the imaged cross. This problem was commonly encountered in the scenes where a broadband light was included. An initial guess promoting the vertical feature by using the ZO (panels B and D) serves sufficiently in this case to retrieve the datacube, irrespective of the mode of the reconstruction itself. On the contrary, an incorrect initial guess cannot be compensated by using the ZO information in the reconstruction routine (see panel

C). It is worth noting that, even for reconstructions ignoring the vertical features, we can attain a reconstruction with low residuals in the detector estimate, i.e., $\left\|I - \widehat{W}D\right\|^2$ from Eq. (1). Therefore, the residuals cannot be generally taken as a good measure to assess reconstruction quality in our system. This was further confirmed by the simulations.

Even though the reconstructed slices of the datacube in Figure 3B and 3D have correct spatial information, the overall spectra are not accurate for the wavelengths below 500 nm, where there was no light intensity due to the use of the OG-515 cutoff filter. Here we observe a significant effect of the regularization weight $\lambda$. By putting stress on the sparsity of the reconstructed signal, i.e., higher $\lambda$ in Equation (1), we obtain a better agreement in the spectrum but the spatial information is impaired. This effect can be observed in Figure 5 presented below. Note that it is possible to improve the reconstruction by restricting it to the range of 500-900 nm. However, the main aim here was to evaluate the limitations of our system and the reconstruction of the imperfect data.

An evaluation of the reconstructions shows that the best results were achieved while using the ZO in both 1) and 2) simultaneously. Hence, we will hereafter show only the comparison between the case of not using the ZO (i.e., standard CASSI approach) and using it both in the initial guess and reconstruction.

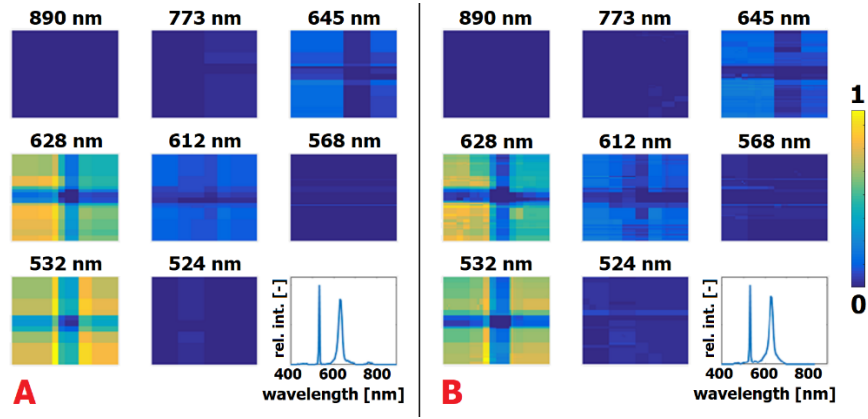

Figure 4. Reconstruction of the scene from Figure 2B (notice that the selected wavelengths are different than in the other figures), each selected spectral slice is normalized to the maximum datacube value, colorbar on the right; (A) Not using ZO, (B) using ZO both in initial guess and operator $\widehat{W}$.

In Figure 4, we can see the reconstructed slices of the scene illuminated by a green laser and a red light-emitting diode. Note that the selected wavelengths of slices shown in Figure 4 are different than in the other figures. The wavelengths were selected to match the maximum spectral intensity of the two peaks. It is possible for the algorithm to distinguish the cross even without the use of ZO, compared to Figure 3A, because the cross is visible in the FO image – see Figure 2B (left), compared to Figure 2D (left). There are only minor differences between the reconstructions in Figure 4A and Figure 4B, and the reconstructed spectrum has a good quality in both cases. Nevertheless, the reconstruction in Figure 4B is slightly superior both in the sense of spatial reconstruction and spectrum quality. Therefore, the spectrally narrow features in the datacube can be well reproduced without the inclusion of the ZO.

Finally, we focused on the scene divided by four color filters illuminated by a broadband light source – see Figure 5. The involvement of the broadband light causes the standard CASSI to face a problem with reconstructing vertical lines in the images, due to spectral shearing. This is highly improved by using the ZO in the reconstruction as can be seen in the borderlines between the quadrants, i.e., the filters, which are visible in Fig. 5B, while in Fig. 5A they are merged together.

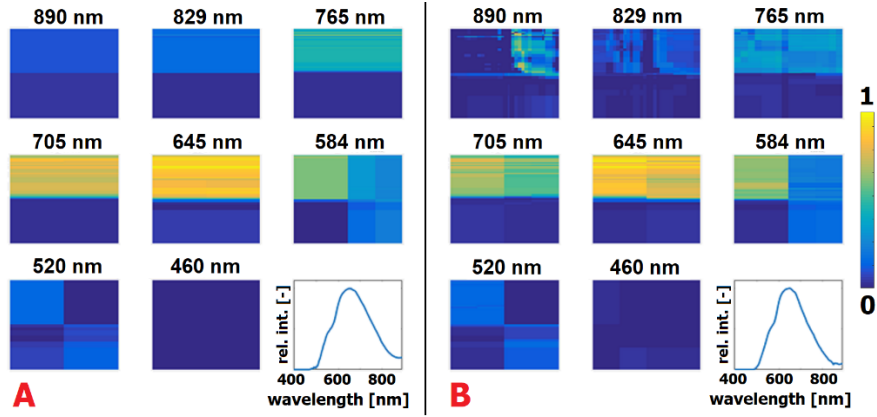

Figure 5. Reconstruction of the scene from Figure 2C, each selected spectral slice is normalized to the maximum datacube value, colorbar on the right; (A) Not using ZO, (B) using ZO both in initial guess and operator $\widehat{W}$.

We extracted the spectrum of the light transmitted through each filter, see Figure 6 B-C, and we compared them with the spectra acquired by a fiber spectrometer (Ocean Optics, Flame), which were corrected for the grating efficiency – see Figure 6A. The colors of the used lines, yellow, red, blue, and green, correspond to the colors of the filters placed in the upper left, upper right, lower left, and lower right quadrant, respectively. Owing to the fact that we used a high $\lambda$ value, we attained spectra which are cropped at 500 nm, in accordance with the used OG515 filter. We attained reasonable agreement between both the reconstructed and the reference spectra. Nevertheless, the reconstruction employing the ZO image reproduces very well even the weak signal from the blue and green filter. The most problematic task is the reconstruction of the overlapping spectra of the red and yellow filter. Here, even the ZO-assisted reconstruction fails to fully reproduce the shape, in spite of reaching a better agreement.

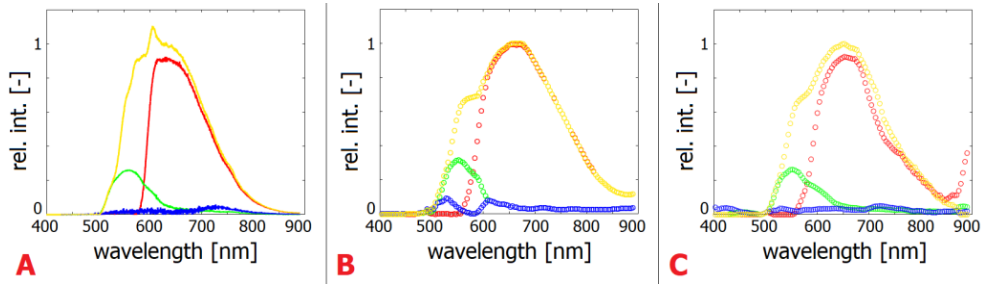

Figure 6: Reconstructed normalized spectra of the individual filters of the scene from Figure 2C, where each line corresponds to a single filter located on the upper left (yellow), upper right (red), lower left (blue), and lower right (green); (A) Independently measured spectra; (B) reconstructed spectra not using ZO; (C) reconstructed spectra using ZO both in initial guess and operator $\widehat{W}$.

We consistently observed that when the ZO is not used in the reconstruction, the resulting spectra are very dependent on the set parameters and it is possible to obtain good results pictured in Figure 6B only with a very narrow set of parameters, while the reconstruction with the ZO (Figure 6C) is much more robust. On top of that, the use of the ZO, in the case of the scene with four color filters, greatly helps to shorten the time needed for the reconstruction (52 seconds without using the ZO vs. 10 seconds using the ZO on a standard laptop) since it improves the initial guess and therefore it converges faster to the results obtained.

To quantify the effect of the ZO usage, we employed the calculations, where we simulated the aberrated detector image and its reconstruction under various conditions. For the sake of comparison between using the ZO and not using it, we evaluated the lowest attainable *difference* – see Eq. (3), between the original and reconstructed datacube.

**Table 1: Difference between the original and reconstructed datacube for different scenes**

|  | Scene A | Scene B | Scene C |
|---|---|---|---|
| Not using the ZO image | 2.1147e-03 | 1.0002e-03 | 9.7144e-04 |
| Using the ZO image | 2.0556e-03 | 9.2579e-04 | 6.0386e-04 |

In Table 1, you can see that the effect of the use of the ZO image depends on the properties of each detected scene. The effect for scenes illuminated with narrow spectral sources is only subtle (Scene A). In the case of broadband light, the influence could be of great importance (Scene C), especially for the scenes, where the spectra are dominated by a single light source. However, for specific scenes and parameter settings, the difference might be lessened (Scene B). Nevertheless, it is worth stressing, that while the level of *difference* might be comparable for both the original and the ZO-assisted CASSI methods, the use of the ZO is much more robust with respect to change in reconstruction parameters.

## 5. Conclusions

We built a broadband hyperspectral single-snapshot camera with a limited number of optical elements based mainly on off-the-shelf optics. Our hyperspectral camera is capable of capturing on a single detector a standard CASSI snapshot of a scene together with a non-dispersed zero-order image. Hence, we can attain more information about the hyperspectral datacube and use the incident light more efficiently.

We carried out hyperspectral imaging on a broad spectral range (500-900 nm) as well as simulations faithfully representing measured data in order to gain more control over the reconstruction algorithm. Due to the inevitable aberrations in the imaging system, we observed that the resulting image highly differed from the ideal case. Therefore, by using the standard CASSI approach, we attained a reliable reconstruction only for simple scenes with quasi-monochromatic light sources.

However, we have achieved a significant improvement in the reconstruction quality by including a ZO image in the CASSI reconstruction. We can employ the ZO image both in the initial guess and the iterative datacube reconstruction. Data show that capturing the ZO image and using it in the reconstruction is beneficial for reconstruction quality and time, which is decreased approx. fivefold. An important factor is that by using the ZO we are able to estimate the datacube in the initial guess very closely to the measured scene.

We observe that the weight of the regularization term in the reconstruction algorithm has a profound effect on the spectral reconstruction quality, where high values of the weight promote correct spectra reconstruction, whereas low values improve the image spatial quality.

In spite of the improvement, the aberrations across the measured broad spectral range still lead to a severe problem with reconstruction quality. However, the results prove that using additional information about the detected scene can partly compensate for the image imperfections. This can be, in the future, utilized in the design of systems for the infrared spectral range, where the reduced imaging system complexity can be of huge importance.

## Funding

## Disclosures

The authors declare no conflicts of interest.